\documentclass{obPhysFluids}
\usepackage{graphics}
\usepackage{obPAPER}
\def\paper#1:#2.#3,#4,#5.{\par\hang\noindent
{\sc\authoryear#1:}\ptitle{#2}\journal{#3}\jvol{#4}\pages{#5}}

\def\book#1:#2.#3,#4.{\par\hang\noindent
{\sc\authoryear#1:}\btitle{#2}\bpubl{#3}\pages{#4}}

\def\proceedings#1:#2.#3,#4.{\par\hang\noindent
{\sc\authoryear#1:}\btitle{#2}\bpubl{#3}\pages{#4}}

\def\authoryear#1,#2,#3#4:{\firstauthor#1,#2:\ifcat#3A\otherauthors#3#4:\else\pyear#3#4:\fi}

\def\otherauthors#1,#2,#3#4:{\ifcat#3A\middleauthor#1,#2:\otherauthors#3#4:\else\lastauthor#1,#2:\pyear#3#4:\fi}

\def\firstauthor#1,#2:{#1, #2}
\def\middleauthor#1,#2:{, #1, #2}
\def\lastauthor#1,#2:{, \& #1, #2}
\def\pyear#1:{\ #1}

\def\ptitle#1{ #1.}
\def\journal#1{ {\it #1,}}
\def\jvol#1{ {\bf #1},}
\def\pages#1{ #1.}

\def\btitle#1{ {\it #1.}}
\def\bpubl#1{ #1,}

\newcommand{\mpl}{\mp}

\title[Statistical mechanics of strong and weak point vortices in a cylinder]
{Statistical mechanics of strong and weak point vortices in a cylinder}
\author[O.\ B\"uhler]{O\ls L\ls I\ls V\ls E\ls R\ns  
B\ls \"U\ls H\ls L\ls E\ls R}
\affiliation{School of Mathematics and Statistics, University of St 
Andrews, \\ St Andrews KY16
9SS, United Kingdom\\[\affilskip]} 
\date{17. Apil 2002}  

\begin{document}

%

\maketitle

\begin{abstract}

The motion of one-hundred point vortices in a circular cylinder is simulated
numerically and compared with theoretical predictions based on statistical
mechanics.  The novel aspect considered here is that the vortices have greatly
different circulation strengths. Specifically, there are four strong vortices
and ninety-six weak vortices, the net circulation in either group is zero, and
the strong circulations are five times larger than the weak circulations.  As
envisaged by Onsager [Nuovo Cimento {\bf 6} (suppl.), 279 (1949)], such an
arrangement leads to a substantial amplification of statistical trends such as
the preferred clustering of the strong vortices in either same-signed or
oppositely-signed pairs, depending on the overall energy level.  To prepare
the ground, this behaviour is illustrated here first by a simple toy model
with exactly solvable statistics.  A microcanonical ensemble based on the
conserved total energy $E$ and angular momentum $M$ for the whole vortex
system is then used, in which the few strong vortices are treated as a
subsystem in contact with a reservoir composed of the many weak vortices.  It
is shown that allowing for the finite size of this reservoir is essential in
order to predict the statistics of the strong vortices accurately.  Notably,
this goes beyond the standard canonical ensemble with positive or negative
temperature.  A certain approximation is then shown to allow a single random
sample of uniformly distributed vortex configurations to be used to predict
the strong vortex statistics for all possible values of $E$ and $M$.  Detailed
predictions for the energy, two-vortex and radial distribution functions of
the strong vortices are then made for comparison with three simulated cases of
near-zero $M$ and low, neutral, or high $E$.  It is found that the statistical
mechanics predictions compare remarkably well with the numerical results,
including a prediction of vortex accumulation at the cylinder wall for low
values of $E$.

\end{abstract}

\newpage

\section{INTRODUCTION}\label{introsec}

The application of statistical mechanics to two-dimensional point vortex
dynamics was first suggested by Onsager$^1$ in a landmark paper in 1949, in
which he sketched a possible explanation for the formation of coherent
vortices on statistical grounds linked to the possibility of negative
(statistical) temperatures for point vortex systems in a bounded domain.
Onsager's suggestions have continued to attract some interest because
two-dimensional fluid dynamics is a relevant paradigm in many applications
such as geophysical fluid dynamics (in which the large-scale quasi-horizontal
flow along stratification surfaces in the atmosphere or oceans is
approximately layerwise two-dimensional), or plasma dynamics under certain
conditions$^2$.  Indeed, a successful application of statistical mechanics to
problems in these fields suggests a hidden degree of predictability that
could easily be obscured by conventional direct numerical simulations.  Such
statistical predictability can be exploited quantitatively at much lower
computational cost than by the brute force simulations of ensembles of many
individual flow realizations.  This is a compelling vision if one considers,
for example, the hugely expensive geophysical climate and weather simulations,
for which currently only a very small ensemble of direct simulations at very
low resolutions is feasible.  Statistical theories based on Onsager's ideas
(and on others that go beyond the point vortex idealization) have already been
used successfully for predicting the detailed behaviour of certain idealized
geophysical flow problems$^3$.  Formidable obstacles remain in order to make
such theories applicable in practice, but the potential rewards are great.

Onsager described his ideas only in qualitative form and the detailed
theoretical exploration of these issues only began with the development of a
mean-field theory by Montgomery and Joyce$^{4,5}$ which has since been
extended and refined mathematically in many ways over the years$^{6-9}$.
%
%
Accurate direct numerical simulations of point vortices over long
times have become feasible over the last two decades$^{10,11}$.
It was suggested$^{11}$ that for geophysical applications the most
relevant values for the number of vortices $N$ lie in an
intermediate range between the regime of low-dimensional chaos
(where $N$ may be less than 10) and the `thermodynamic' regime in
which $N\to\infty$ in some subtle limit.  This point of view is
also taken here, where only cases with $N=100$ are studied.

Now, the topic of the present paper goes back to a suggestive remark made by
Onsager concerning the characteristic appearance of vortex distributions in a
negative temperature state, and which to my knowledge has not been considered
explicitly since.  In his masterful succinct exposition, Onsager says that in
such a state 

``\ldots vortices of the same sign will tend to cluster,---{\it
preferably the strongest ones\/}---, so as to use up excess energy at the
least possible cost in term of degrees of freedom \ldots the weaker vortices,
free to roam practically at random, will yield rather erratic and disorganised
contributions to the flow'' (my italics).  

This encapsulates two crucial insights: (a) that strong vortices will be more
predictable than weak ones; and (b) that the maximum disorder of the flow as a
whole (which is implied by the statistical theory) will be achieved in a
remarkably inhomogeneous, composite manner, in which the relative order of the
strong vortices will be more than compensated for by the increased disorder of
the weak ones.  The present paper is an attempt to verify and exploit both of
these insights in the simplest possible setting that allows comprehensive
numerical and theoretical exploration.

To the best of my knowledge, detailed previous studies have all
focused on the case of identical (or nearly identical$^{11}$) absolute
vortex circulations, or on the even more restricted case of
identical vortex circulations throughout.  In these cases, the
statistical behaviour of the flow is in some sense directly
determined by the constraint of fixed total energy: large energy
values must require the coming together of same-signed vortices,
and vice versa for low energy values (see \ref{hamsec} below).
In other words, the occurrence of vortex clusters is then
enforced directly by the total energy constraint.  However, once
different vortex strengths are present there is scope for
interestingly different behaviour, in which some vortices cluster
spontaneously whilst others do not.

The plan for the paper is as follows: \S\ref{hamsec} introduces the main
features of the studied Hamiltonian point vortex system and corrects some
minor errors in related works; \S\ref{switsec} briefly illustrates Onsager's
suggestion by an analogy with a simple toy model, which serves to prepare to
ground for the vortex theory; \S\ref{numsec} presents direct numerical
simulations of the vortices; \S\ref{smsec} derives statistical mechanics
predictions and applies these to the simulations; and some concluding remarks
are given in \S\ref{concsec}.

\section{HAMILTONIAN POINT VORTEX DYNAMICS}\label{hamsec}

The Hamiltonian form of the equations of motion for $N$
point vortices with circulations $\Gamma_i$ and instantaneous Cartesian
coordinates $\bx_i(t)=(x_i,y_i)$ (where $i = 1,2,\ldots,N$) are:  
\beql{HamFormEOM}
\Gamma_i\,\dert{x_i} = + \pder{H}{y_i},\qquad \Gamma_i\,\dert{y_i} = -
\pder{H}{x_i}.
\eeq
The pairs $(x_i,y_i)$ are canonical phase space coordinates, with
invariant phase space volume element $\dbx_1\dbx_2\ldots\dbx_N$.
In the special case of a circular cylinder with radius $R$
centred at the coordinate origin$^6$ the wall boundary condition
is satisfied$^{12}$ by placing for each physical vortex at
location $\bx$ a single image vortex with opposite circulation at
location $\bx R^2/|\bx|^2$.  This leads to an invariant
Hamiltonian $H(\bx_1,\ldots,\bx_N)$ as
\beqal{HFCC}
H  = -\frac{1}{4\pi} \sum_{i>j}^{N,N} 
\Gamma_i\Gamma_j\, \ln(r_{ij}^2) &+& \frac{1}{4\pi} \sum_{i=1}^N
\Gamma_i^2\,\ln(R^2- r_{i}^2) \\
&+& \frac{1}{4\pi} \sum_{i>j}^{N,N} 
\Gamma_i\Gamma_j\,\ln(R^4-2R^2\bx_i\cdot\bx_j +
r_i^2r_j^2)\nonumber,
\eeqa
where $r_{i}^2 = x_i^2 + y_i^2$, $r_{ij}^2 = (x_i-x_j)^2 + (y_i-y_j)^2$, and
the asymmetry in the logarithmic terms arises because the image vortices do
not move with the implied physical velocity at their location.  In other
words, the circular cylinder wall cannot be removed.  The double sums run over
all pairs $i>j$, i.e.\ there are $N(N-1)/2$ individual terms in these.  As
$0\le r_i\le R$, the phase space is clearly bounded and its volume is $(\pi
R^2)^N$.  In the generic case of $\Gamma_i$ with different signs we clearly
have $H\in(-\infty,+\infty)$ due to various combinations of terms with $r_i\to
R$ or $r_{ij}\to0$.  Indeed, even at fixed $H=E$, say, individual terms in
\Ref{HFCC} can go to $\plm\infty$ whilst adding up to a finite number.  

The first sum in \Ref{HFCC} involves the usual free-space interaction term,
which goes to $+\infty$ as $r_{ij}\to0$ for same-signed vortex pairs, and vice
versa for oppositely-signed dipoles.  As is well known, this symmetric
appearance masks a quite distinct dynamical behaviour in these two cases, with
same-signed vortices orbiting each other whilst oppositely-signed ones
propagating along a straight line.  The second sum involves a self-interaction
term for each vortex, which leads to counter-clockwise propagation at fixed
$r_i$ for $\Gamma_i>0$ and vice versa for $\Gamma_i<0$.  Unlike the pair
interaction terms, this term always goes to $-\infty$ as $r_i\to R$, which is
linked to the ever-closer approach of the vortex to its oppositely-signed
image in this limit.  In other words, the cylinder wall is a location of
infinite negative energy for each vortex.  Finally, the third sum in
\Ref{HFCC} involves the interaction of each vortex with the images of all
other vortices.  Its terms become singular only if $r_i,r_j\to R$ and
$r_{ij}\to0$.

One can note in passing that the dynamically active nature of the
cylinder wall that is expressed by the self-interaction terms
distinguishes the present case from the previously studied
doubly periodic case$^{10,11}$.  These self-interaction terms add
advection parallel to the wall to the dynamics, which somewhat
enhances the mobility of individual vortices.  Also, one can note
that in the cylinder case propagating vortex dipoles split up
when they approach the cylinder wall and then propagate along the
wall into opposite directions.  Unless they collide with other
vortices beforehand, the vortices would then rejoin on the other
side of the cylinder, and again enter the interior as a dipole.
Such vortex behaviour might in fact be relevant for vortex
dynamics on beaches, where the vicinity of the shoreline has a
similar effect as the cylinder wall$^{13}$.

%
%

The infinities of the various terms in \Ref{HFCC} occur on a set
of measure zero in phase space volume, but they nevertheless have
an impact on direct numerical simulations as well as on
statistical theories, especially those with non-uniform
phase-space measures (e.g.\ \Ref{Zdef} below).  Indeed, the
possibility of negatively infinite self-interaction energy is
important even in the simplest case $\Gamma_i=\Gamma=$~const,
although this seems to have been overlooked at times.  For
instance, Caglioti {\it et al\/}.$^{8}$ consider the conditions
for existence of the usual canonical partition function $Z$
defined by the total phase space integral
\beql{Zdef}
Z = \int \exp(-\beta H)\,\bdx^N,
\eeq
where $\beta$ is the usual parameter inversely proportional to the
(statistical) temperature.  They state that $Z<+\infty$ exists for fixed
$N$ and $\Gamma$ if and only
if 
\beql{betarange1}
\beta\in\left(\frac{-8\pi}{\Gamma^2 N},+\infty\right),
\eeq
where the finite negative range is needed in order to bound the
importance of same-signed vortex collisions with their infinite
positive energies.  However, this miscalculates the importance of
the {\it negative\/} infinite energies as vortices approach the
wall.  Indeed, $Z$ for a single vortex in a cylinder is easily
evaluated as
\beql{Zsingle}
Z = \frac{\pi}{1-\beta_*} R^{2(1-\beta_*)}\quad\mbox{only if}\quad \beta_*=
\frac{\beta\Gamma^2}{4\pi} < 1,
\eeq
otherwise $Z<+\infty$ does not exist.  It seems that this implies that
\Ref{betarange1} in Caglioti~{\it et al.} should be replaced by
the slightly more symmetrical
\beql{betarange2}
\beta\in\left(\frac{-8\pi}{\Gamma^2 N},\frac{+4\pi}{\Gamma^2}\right),
\eeq
which exhibits a finite temperature range also for $\beta>0$.
(In the particular asymptotic limit subsequently studied in that
paper$^8$, the re-scaled upper limit in \Ref{betarange2} still
tends to $+\infty$, so the subsequent results may well remain
intact.)

Somewhat surprisingly, this means that a point vortex system with
$\Gamma_i=\Gamma$ and bounded by a solid wall, if coupled to an infinite
energy reservoir at positive temperature $T$, will collapse to the wall as $T$
drops towards $\Gamma^2/4\pi$, where Boltzmann's constant has been set to
unity.  Some evidence for such behaviour is given in \S\ref{smsec} below.  Of
course, the point vortex model will lose its physical significance when this
happens, i.e.\ the finite core and finite self-energy of physical vortices
will become important in this limit.  Also, any coupling to a {\em finite\/}
energy reservoir will arrest the collapse (cf.\ \S\ref{switsec}).

Now, in the remainder of this paper the following set-up will be studied.  The
total number of vortices $N=100$ will be split into $N_A=4$ strong vortices
and $N_B=96$ weak vortices with respective circulations 
\beql{circvals}
\Gamma_A=\plm10\pi\quad\mbox{and}\quad \Gamma_B=\plm2\pi.
\eeq
The net circulation in either group is zero.  This is a natural constraint for
physical situations that have arisen from localized vortex forcing, which
always produces vortex dipoles with zero net circulation$^{13}$.  These
particular values have been chosen in order to focus on the most complex
scenario, as follows.

Because absolute energy values in \Ref{HFCC} are meaningless (unlike in most
classical systems), the relevant measure of importance of the individual sums
in \Ref{HFCC} comes from considering their variance over the phase space.
Neglecting $\Gamma_A\Gamma_B$ interaction terms between strong and weak
vortices, it turns out (cf.\ \S\ref{switsec} below) that the total variance of
the $\Gamma_A^2$ terms in the double sums scales approximately as
$O(N^2_A\Gamma_A^4)$, and correspondingly so for the weak vortices.  Therefore
the parameters have been arranged such that these variances are approximately
equal.  The total variance of the $\Gamma_A^2$ self-interaction terms in
\Ref{HFCC} scales as $O(N_A\Gamma_A^4)$ and it turns out that for the chosen
set-up the numerical pre-factor makes this term significant in size compared
with the term $O(N^2_A\Gamma_A^4)$.  However, for the weak vortices the
self-interaction variance $O(N_B\Gamma_B^4)$ is small compared to the term
$O(N^2_B\Gamma_B^4)$.  In summary, strong and weak vortices are expected to
interact vigorously and the dynamics of the strong vortices is in addition
significantly affected by the presence of the wall.

It remains to consider a  second invariant that arises due to  the azimuthal
symmetry of the Hamiltonian $H$ in \Ref{HFCC}, namely
the invariant angular momentum 
\beql{AngMomDef}
\hat M = \frac{1}{2\pi}\sum_{i=1}^N \Gamma_i\, r_i^2.
\eeq
There are no other invariants, so vortex motions with $N>2$ are presumably
non-integrable in the cylinder.  Unlike $H$, the invariance of $\hat M$ is not
robust in the sense that a small disturbance of the problem (say, perturbing
the cylinder wall to be elliptical) would destroy the invariance of $\hat M$.
Nevertheless, in the present case $\hat M$ clearly plays a r\^ole and needs to
be considered formally on the same footing as $H$.  It is straightforward to
show that for the chosen set-up the strong and weak vortex contributions to
the variance of $\hat M$ are again roughly equal.  

Finally, it is noteworthy that in a set-up in which all the $\Gamma_i$ are
sign-definite, the conservation of $\hat M$ implies that the accessible phase
space is bounded even without a cylinder wall, and this has been used to study
negative temperature states of such a set-up using only the first term in
\Ref{HFCC}.  However, in the present set-up with $\Gamma_i$ of either sign
this cannot be done (contrary to assertions sometimes made$^{7,14}$, where
\Ref{AngMomDef} was misquoted with $\Gamma_i$ replaced by $\Gamma_i^2$).

%
%
\section{A SOLVABLE TOY MODEL}\label{switsec}

Here a toy model with exactly solvable statistical mechanics is discussed in
order to prepare the ground for the statistical mechanics of the vortex
system.  The toy model is
the one-dimensional Ising model without external magnetic field, which can be
thought of as an assembly of independent switches.  Let there be $M$ switches
and let each switch either be in an ``up'' or ``down'' position, with
corresponding energy values $E_i=\plm\epsi$, where $i=1,2,\ldots,M$ and the
constants $\epsi$ describe the individual strength of the switches.  The
finite-sized discrete state space of the system is formed by the $2^M$
different states of all the switches.  The total energy is
\beql{Ettot}
E = \sum_{i=1}^{M}\, E_i\quad\mbox{with range}\quad - \sum_{i=1}^{M}\, 
|\epsi| \leq E
\leq + \sum_{i=1}^{M}\, |\epsi|.
\eeq
The energy extrema correspond to exactly one state each, and it is clear that
there are only a few states with energies near these extrema.  This scarcity
of states is of crucial importance for the statistical mechanics of this
system.  Now, in an analogy with vortex dynamics each of the $M$ switches
corresponds to one of the $O(N^2)$ interaction terms in the first sum in
\Ref{HFCC}, the other sums being disregarded.  Each $\epsi$ then corresponds
to a pairing $\Gamma_i\Gamma_j$, and the up/down switch positions correspond
crudely to the variability of the logarithms.  Surprisingly, it will turn out
that this crude analogy can already explain a number of features of the vortex
system.

\subsection{Canonical statistical mechanics}

We first imagine the system in contact with an infinite energy reservoir at
temperature $T$ and inverse temperature $\beta=1/T$.  The individual switch
statistics are then independent from each other and hence it suffices to
consider the $i$th switch in isolation.  (The analogous statement is of course
not true in the vortex system, which is a coupled $N$-body problem.)  The
probabilities for the switch to be in the up/down position are then equal to
$\exp(\mpl\beta\epsi)/Z_i$ respectively, where the normalization constant is
the partition function $Z_i=2\cosh(\beta\epsi)$.  The resultant average switch
energy $U_i$ is
\beql{Utswitch}
U_i = -\epsi\tanh(\beta \epsi),
\eeq
which shows that there is no energy equipartition between the switches unless
$\beta=0$, and that regardless of the sign of $\epsi$ a positive temperature
corresponds to negative $U_i$ and {\it vice versa}.  The behaviour of the
switch as a function of $\beta\epsi$ is easily characterized: if
$\beta|\epsi|\gg1$ (i.e.\ $T\to0+$) we have $U_i\approx -|\epsi|$ and the
switches are increasingly locked into their low-energy positions.  On the
other hand, if $\beta|\epsi|\to 0+$ (i.e.\ $T\to\infty$) the switches become
increasingly disordered and $U_i$ goes to zero.  A corresponding scenario
unfolds for $\beta<0$, which shows that positive and negative temperature
statistics are perfectly symmetric here.  All this applies qualitatively to
the vortex system as well, though the near-symmetry between positive and
temperature states is lost there when considering a mean-field theory$^{4,5}$.

The switch fluctuations can be analyzed by considering the number
$U_i/\epsi=-\tanh(\beta \epsi)$, which is the average switch
position.  (Identical conclusions are reached by considering the
variance of the switch energy, which is equal to $-\partial
U_i/\partial \beta=\epsi^2/\cosh^2(\beta\epsi)$.)  Absolute
values of this number near unity mark ordered, predictable
behaviour, whereas absolute values near zero mark disorder and
randomness.  For uniform $\epsi$ order increases as $|\beta|$
increases.  Consider now the case of there being two distinct
types of switches: strong and weak, respectively, such that
$|\eps_A|>|\eps_B|$, in obvious analogy to the vortex set-up.
Clearly, for the same value of $\beta$ the behaviour of the
strong switches is going to be more ordered than that of the weak
switches, i.e.\ $|U_A/\eps_A| \geq |U_B/\eps_B|$, with equality
holding only when $\beta=0$.  Indeed, if $|\eps_A|\gg|\eps_B|$
then the strong switches can exhibit significantly more order
than the weak switches, which illustrates the first part of
Onsager's remark.

This point about order/disorder can be made in another way by looking at the
contributions to the entropy of the system that are made by the strong and weak
switches, respectively.  The usual expression for the entropy in the canonical
formalism is
\beql{Stswitches}
S_i = \ln Z_i + \beta U_i = \ln(2\cosh(\beta\epsi)) -
\beta\epsi\tanh(\beta\epsi).
\eeq
The switch entropy $S_i(\beta\epsi)$ is hence a positive even
function of its argument with
maximum $S_i(0)=\ln(2)$ at the origin, and with monotone decrease to zero with
increasing $|\beta\epsi|$.  (At $\beta=0$ and $\beta\to\plm\infty$ the value
of $S_i$ is compatible with the microcanonical definition of entropy as the
logarithm of the number of permissible states.)  Again, this means that at
fixed $\beta$ we have $S_A\leq S_B$, i.e.\ the strong switches contribute less
to the entropy than the weak switches.

A dynamical interpretation of this remarkably composite, inhomogeneous entropy
distribution is suggested by the alternative variational formulation of
canonical statistical mechanics in terms of maximum (information) entropy at
fixed mean energy.  
In this view, the strong switches tend to become more ordered,
and hence contribute less to the entropy, because in doing so
they absorb the right amount of energy to allow the weak switches
to be as disordered as possible.  This peculiar sharing-out of
the energy leads a distribution of maximum total entropy, which
then emerges as a truly composite, interactive feature of the
system.  This illustrates the second part of Onsager's remark.

\subsection{Microcanonical statistical mechanics}

We now turn to consider the microcanonical statistics of the toy model, in
which the total energy is fixed at a certain value $E$ such that only states
with this energy value are permitted and all such states are then deemed
equally likely.  This is the proper setting in which to analyze numerical
simulations at conserved energy under an ergodic approximation, and the
formalism developed here will be directly relevant to the vortex dynamics
discussed later.  

The total energy constraint now couples the individual switch statistics.
Specifically, a subset of switches now behaves like a subsystem in contact
with a {\em finite\/} energy reservoir formed by the other switches, and the
key question is to analyze the statistical mechanics of such a subsystem.  To
this end we split the $M$ switches into a subsystem $A$ and a ``reservoir''
$B$ such that $M_A+M_B=M$ and $M_B\gg1$, assuming from now on that $M\gg1$ to
begin with.  We have
\beq
E = E_A + E_B
\eeq
for all permissible states.  The probability of a particular subsystem 
state with energy $E_A$ is then proportional to the number of 
reservoir states with $E_B=E-E_A$.  Now, the density of 
reservoir states per unit energy interval is that of a sum of 
$M_B\gg1$ independent zero-mean random functions with finite variances, 
which by the central limit theorem can be approximated by the 
continuous function 
\beql{ntres}
p_0(E_B) = 
\frac{1}{\sigma_B\sqrt{2\pi}}\exp\left(\frac{-E_B^2}{2\sigma_B^2}\right),
\quad\mbox{where}\quad \sigma_B = \sum_{i=1}^{M_B}\,\epsi^2
\eeq
is the energy variance of the reservoir.  The subscript $0$ denotes that this
probability corresponds to a uniform  distribution over all possible
states.  For a particular subsystem state with $E_A$ we therefore obtain the
probability
\beqal{probtA}
\mbox{\sl Prob\/}\left\{\mbox{subsystem state with $E_A$}\right\} &\propto&  
p_0(E_B=E-E_A) \propto
\exp\left(\frac{-(E-E_A)^2}{2\sigma_B^2}\right)\nonumber \\ 
&\propto& \exp\left(\frac{E\,E_A}{\sigma_B^2} 
-\frac{E_A^2}{2\sigma^2_B}\right),
\eeqa
where a factor independent of $E_A$ has been absorbed into the proportionality
constant determined by normalization.  This means that for small subsystem
energies $E^2_A\ll2\sigma^2_B$ the relative probabilities are described by
canonical statistics with an inverse temperature $\tilde\beta=-E/\sigma_B^2$.
We note in passing that if $|E_A|\ll|E_B|$ then $\tilde \beta$ agrees with the
microcanonical definition of inverse temperature of the reservoir as the
derivative of $\ln p_0(E_B)$ with respect to $E_B$.

Now, for larger subsystem energies the finite size of the
reservoir is felt via the second term in \Ref{probtA}, which
decreases the likelihood of such states and marks the departure
from canonical statistical mechanics for the subsytem.
(Coincidentally, the statistics of a {\em single\/} switch in
contact with the finite reservoir are precisely canonical, due to
the symmetry of its two energy levels.)  Also, one can note that
the probability distribution \Ref{probtA} admits a somewhat
unusual maximum-entropy formulation in which the subsystem
information entropy is extremalized subject to {\em two\/}
constraints, namely that of fixed average energy as well as fixed
energy variance.  It is straightforward to show that such a
procedure leads to a probability $\propto\exp(-\tilde\beta E_A -
\tilde\gamma E_A^2)$ for a subsystem state with energy $E_A$, in
accordance with the structure found in \Ref{probtA}.

The above shows how microcanonical statistical mechanics can be used to
calculate the statistics of a subsystem in contact with a finite reservoir,
even beyond the usual asymptotic limit $M_B\to\infty$, $E/M_B=O(1)$ in which
canonical subsystem statistics would emerge.  The important reservoir energy
variance $\sigma^2_B$ in \Ref{ntres} enters the definition of $\tilde \beta$
and it also demarcates the finite size of the reservoir as felt by the
subsystem.  In analogy with the toy model and \Ref{ntres}, the reservoir
variance in the vortex system is proportional to $N_B^2\Gamma_B^4$, as was
previously asserted.

%
%
\section{NUMERICAL SIMULATIONS}\label{numsec}

Direct numerical simulations of the vortex system set-up from \S\ref{hamsec}
are described and analyzed.   The comparison with statistical mechanics
predictions follows in \S\ref{smsec}.

\subsection{Numerical model details}

The model integrates the dynamical equations derived from \Ref{HamFormEOM} and
\Ref{HFCC}, i.e.
\beqal{EoM}
\dert{\bx_i} &=& \sum_{j=1,j\neq i}^N\, 
\frac{\Gamma_j/2\pi}{(\bx_i-\bx_j)^2}\, (y_j-y_i,x_i-x_j)\\
&+& \sum_{j=1}^N\, 
\frac{\Gamma'_j/2\pi}{(\bx_i-\bx'_j)^2}\, (y'_j-y_i,x_i-x'_j)\nonumber
\eeqa
As noted in \S\ref{hamsec}, each vortex has an image vortex with parameters
\beql{ImVortDef}
\Gamma'_j = -\Gamma_j,\quad \bx'_j = \bx_j\frac{R^2}{x_j^2+y_j^2}
\eeq
to satisfy the wall boundary condition.  Notably, the second sum in \Ref{EoM}
includes a self-interaction term with $j=i$.  Strong and weak vortices were
chosen as in \Ref{circvals} and for definiteness the cylinder radius $R=5$.
By re-scaling \Ref{EoM} the present simulations can be mapped onto simulations
with arbitrary finite $R$ and $\Gamma_B$ provided the ratio
$|\Gamma_A/\Gamma_B|=5$ remains the same.

The numerical model itself is a standard Runge--Kutta scheme of 4/5th order
with adaptive timestep refinement.  The integrations are performed with double
precision and in cycles of $\delta t=0.01$, using adaptive timestep refinement
until the error is less than the tolerance set to $10^{-9}$.  No
regularization of the equations for numerical purposes was needed, e.g.\ there
was no near-field smoothing to prevent large velocities, and the unusually low
error tolerance is required to integrate safely through intermittent episodes
in which vortices are getting very close to each other, or to the wall.  These
intermittent episodes lead to a model performance that can vary by a factor of
ten over a run.  The total energy and angular momentum are conserved with very
high precision and all runs were performed on a single-processor workstation.

\subsection{Description of three model runs}

Under the ergodic approximation the statistics of a run are determined by the
invariant values of its Hamiltonian $H=E$ and its angular momentum $\hat M=M$,
say.  All runs had values of $M$ very close to zero.  This is because no
exploration of the r\^ole of $M$ was intended because of the non-robust nature
of this invariant noted in \S\ref{hamsec}.  Three different energy values
(denoted Low, Neutral, and High) were chosen to give runs broadly
corresponding to regimes of positive, zero, and negative microcanonical
temperature.  The specific values for the three runs were
\beql{EMruns}
E = \{-197, 221, 628\},\quad M = \{2.1, 4.1, 2.3\}
\eeq
and they were determined as follows$^{10}$.  A random population
of $10^5$ vortex configurations was generated in which each of the hundred
vortices was placed independently with uniform probability anywhere inside the
cylinder.  The $H$ values were then computed from \Ref{HFCC} (this being the
computationally expensive step) and a histogram was formed to give the
probability density function~(pdf) for the total energy $p_0(E)$, as plotted
in \fref{fHamAm}a.  The average energy is non-zero because of the
sign-definite effect of the self-interaction terms in \Ref{HFCC} and also
because there are $O(N)$ more oppositely-signed terms than same-signed terms
in the double sums.
\epsherel{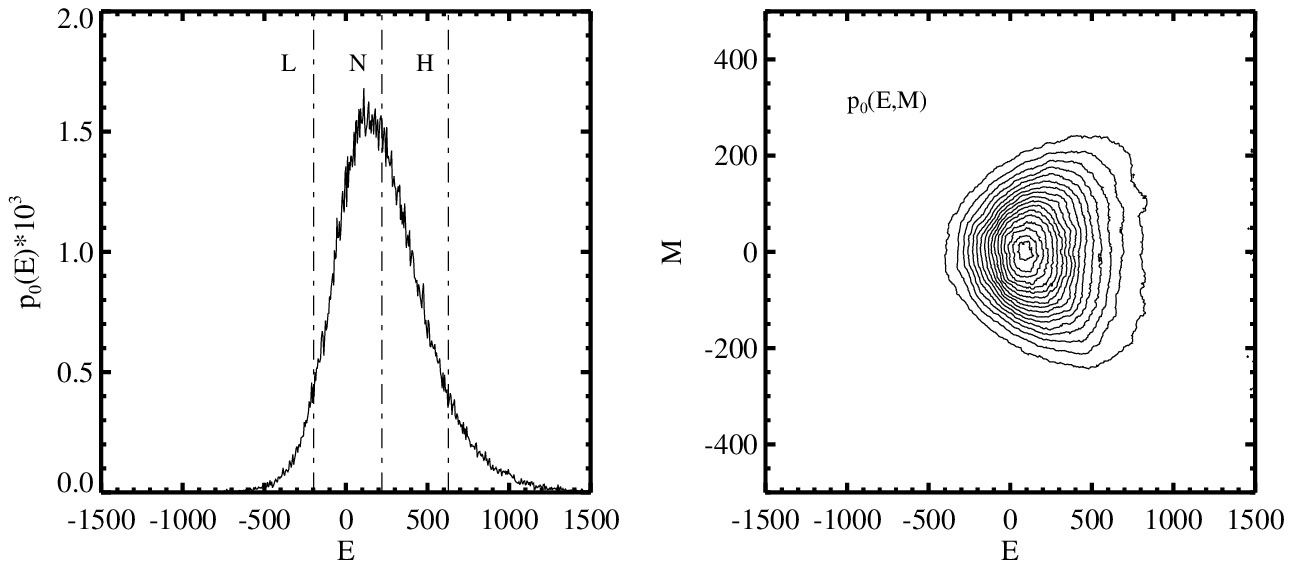}{(a) Total energy pdf (based on $10^5$
random samples with uniform distribution) with indications of the total energy
levels for the Low, Neutral, and High runs.  The average total energy is $222$
with variance $284^2$.  (b) Joint total energy--angular momentum pdf, after
some smoothing.  The correlation coefficient between $E$ and $M$ is
$10^{-3}$.}{fHamAm}
The corresponding pdf for the total angular momentum (not
plotted) is close to a zero-mean normal distribution with
variance equal to $\Gamma_i^2R^4/48\pi^2$ summed over all
vortices.  Also plotted in \fref{fHamAm}b is the joint pdf of
total energy and total angular momentum $p_0(E,M)$, which
indicates approximate statistical independence of $E$ and $M$.

At the begin of each run the four strong vortices were placed in the same
positions, i.e.\ these vortices were symmetrically spaced in azimuthal angle
at a common radius $r=3$, with alternating circulation from vortex to vortex.
This is very close to an exact steady state of four vortices in a cylinder,
which occurs at radius $r=(\sqrt{17}-4)^{1/4}R\approx2.96$, and hence the
strong vortices are essentially set into motion by the weak vortices.  The
positions of the weak vortices were taken from uniformly distributed random
samples that were generated until a sample with suitable $E$ and $M$ was
found.  The resulting initial states are displayed in \fref{fVinits}.
\epsherel{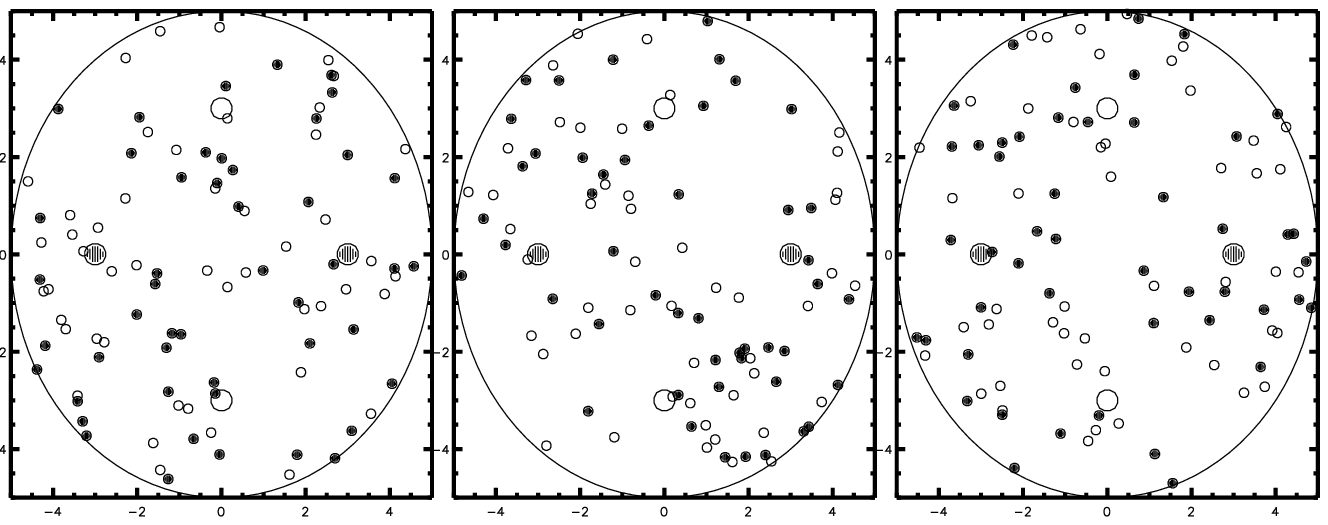}{Inital states for the three runs with Low, Neutral,
and High energy levels, from left to right.  Large circles show the strong
vortices, and black or white colour indicates positive or negative
circulation.}{fVinits}
It is probably fair to say that it is impossible to guess by inspection of
\fref{fVinits} whether and in what way the long-term statistics of the strong
vortices will differ in these runs.

All runs were now integrated up to $t=750$, which corresponds to
many hundred cylinder traversals of each vortex, and $3000$
instantaneous states of the evolution were stored at time
intervals of $\Delta t=0.25$.  Very different behaviour of the
strong vortices was observed.  The low-energy run showed a strong
tendency of vortices sticking close to the wall, or forming
short-lived vortex dipoles.  The high-energy run showed a strong
tendency to form same-signed vortex pairs that persisted a
comparatively long time.  The neutral-energy run showed a mixture
of both behaviours.  In all cases there were fluctuations around
this behaviour, but the self-organization of the vortices into
these typical patterns was conspicuously clear.
%
%
%

Several quantitative diagnostics for the strong vortices were
computed: their energy $E_A$ (i.e.\ \Ref{HFCC} evaluated using
only the strong vortices), the distance between same-signed
vortex pairs, and the distance between oppositely-signed vortex
dipoles.  The mean values and standard deviations of these quantities
over the duration of the simulations are summarized in
table~\ref{tmom}, with pdfs presented in the next section.
\begin{table} 
\begin{center} 
\begin{tabular}{lclll} 
& & Low & Neutral & High                        \\ 
\hline 
Energy & $E_A$    & -23 $\plm$ 126    &  141 $\plm$ 111 & 269 $\plm$ 136  \\ 
{\sl predicted\/} &       & -64 $\plm$ 136    &  129 $\plm$ 117 & 242 $\plm$
152
\\[.5ex]
Same-sign distance &$r_{ij}$ & 5.1 $\plm$ 2.1    &  4.5 $\plm$ 1.9 & 3.7 $\plm$ 2.0\\ 
%
{\sl predicted\/} &    & 5.1 $\plm$ 2.1    &  4.5 $\plm$ 1.9 & 3.9$\plm$ 2.1\\[.5ex]  
Opp.-sign distance &$r_{ij}$& 4.5 $\plm$ 2.5  &  4.4 $\plm$ 2.1 & 4.6 $\plm$ 1.8\\ 
%
{\sl predicted\/} &       & 4.6 $\plm$ 2.5  &  4.5 $\plm$ 2.1 & 4.5 $\plm$  1.9\\[.5ex]
Radius &$r_{i}$   & 3.6 $\plm$ 1.2    &  3.3 $\plm$ 1.1 & 3.1 $\plm$ 1.0\\ 
{\sl predicted\/} &    & 3.6 $\plm$ 1.2    &  3.3 $\plm$ 1.1 & 3.1$\plm$ 1.1\\[.5ex]  
\end{tabular} 
\end{center} 
\caption{Mean values and standard deviations over the duration of
the simulation vs.\ theoretical 
predictions for various quantities.  For the distances $r_{ij}$ and $r_i$
these quantities are averaged over all relevant combinations of
$i,j\in\{1,2,3,4\}$ if $\Gamma_{1,2}=-10\pi$ and $\Gamma_{3,4}=+10\pi$.  }
\label{tmom} 
\end{table} 
These numbers make clear that significant statistical differences are indeed
observed between the runs.  With increasing overall energy $E$ the average of
the subsystem energy $E_A$ (which initially has equal value in all runs)
increases and the average distance between same-signed vortex pairs decreases.
Also, at the neutral energy level (with zero microcanonical temperature) there
is no preference between pairing of same-signed or oppositely-signed vortices.
The average distance between oppositely-signed vortex
dipoles remains roughly constant as $E$ is increased and close to a value
corresponding to uniformly random placement of the vortices, which is $4.5$
(with standard deviation $2.1$).    However, the standard deviation of this
quantity reduces with $E$.  

Table~\ref{tmom} also includes predictions based on the theory described in
\S\ref{smsec} below.  These predictions are generally quite accurate, except
for $E_A$ in the low-energy case and for the same-signed $r_{ij}$ in the
high-energy case.  Sample autocorrelation functions based on the
numerical time series feeding into these averages were computed in order to
estimate confidence intervals.  These suggested long-lived oscillations in the
low-energy $E_A$ with periods of about $t\sim50$ as well as a slowly decaying
autocorrelation in the high-energy $r_{ij}$, which presumably is linked to
long-lived vortex pairs in that case.  Confidence intervals based on these
estimates put the observed discrepancies at the borderline of statistical
significance.

%
%
\section{COMPARISON WITH STATISTICAL MECHANICS}\label{smsec}

The numerical results are compared with pdfs estimated based on microcanonical
statistical mechanics for the whole system.  The general estimation procedure
is described in some detail, not least because it contains an important {\sl
ad hoc\/} approximation that significantly reduces computational cost.

\subsection{Estimation of pdfs}

The estimation of pdfs is based on phase space densities
$\rho(\bx_1,\ldots,\bx_N)$, which for convenience in the present case are
normalized such that
\beql{rhonorm}
\int \rho\,\dbx^N = 1,
\eeq
where the integral is extended over the whole phase space.  Relative to a
chosen $\rho$ the pdf of any phase space function $\Phi(\bx_1,\ldots,\bx_N)$
taking real values $\phi$ is defined as
\beql{pofphidef}
p(\phi) = \int \delta(\Phi-\phi)\rho\,\dbx^N =
\int_{\Phi=\phi}\frac{\rho\mbox{dA}} {|\grad\Phi|}.
\eeq
The scaling properties of the Dirac-$\delta$ function succinctly capture the
thickness of the layers $\Phi\in[\phi,\phi+d\phi]$ measured in the surface
integral on the right.  Multiple pdfs $p(\phi_1,\phi_2)$ are defined
analogously using products of $\delta$-functions.  Pdfs defined by
\Ref{pofphidef} can be estimated numerically by forming histograms of
$\Phi(\bx)$ based on random samples of $\bx$.  In theory, optimal convergence
of such a procedure requires importance sampling, in which $\rho$ is used as
the density of the random sample.  In practice, using a uniform density
coupled with histogram increments proportional to $\rho$ is much cheaper in
the present case of a finite phase space.  The special case of a uniform
density $\rho_0=(\pi R^2)^{-N}$ is particularly important, for example
$p_0(E)$ in \fref{fHamAm} has been estimated from
\beql{p0Edef}
p_0(E) = (\pi R^2)^{-N} \int \delta(H-E)\,\dbx^N.
\eeq
All pdfs calculated from the uniform density are denoted by $p_0(\cdot)$.

The usual microcanonical density based on energy $H\in[E,E+dE]$ is defined as 
\beql{rhomcdefold}
\rho_E = \frac{\delta(H-E)}{\int \delta(H-E)\,\dbx^N}
\eeq
and corresponding pdfs will be denoted by $p_E(\cdot)$.
For simplicity, consideration of the angular momentum invariant is deferred until
later.  From \Ref{pofphidef} one obtains
\beql{pmcdefold}
p_E(\phi) \propto \int \delta(\Phi-\phi)\delta(H-E) \,\dbx^N \propto
p_0(\phi,E),
\eeq
up to an overall normalization factor.  This means that $p_E(\phi)$ can in
principle be evaluated from a joint pdf $p_0(\phi,E)$ based on the uniform
distribution.  However, for a particular value of $E$ this is computationally
very expensive, as most samples have to be discarded.  On the other hand, if
$\Phi$ depends only on a subset of the variables then its pdf can be much
simplified, as follows.  

Specifically, let $\bx_A$ and $\bx_B$ denote the coordinates of all the strong
and weak vortices, respectively.  Then $\dbx_A\dbx_B=\dbx^N$ and we consider
only functions $\Phi(\bx_A)$ from now on.  From \Ref{pmcdefold}
one obtains
\beql{pmcdef}
p_E(\phi) \propto \int \delta(\Phi-\phi) \left[\int \delta(H-E)
\,\dbx_B\right]\,\dbx_A \propto \int \delta(\Phi-\phi) \rho_E(\bx_A) \,\dbx_A,
\eeq
where the integral in squared brackets is the marginal density
$\rho_E(\bx_A)$.  Now, if it were the case that $H=H_A(\bx_A)+H_B(\bx_B)$ (as
was true
in the toy model) then
\beqal{sortos}
\rho_E(\bx_A) &\propto& \int \delta(H_B+H_A-E) \,\dbx_B \propto
p_0(E_B=E-H_A),\quad\mbox{where}\\
p_0(E_B) &\propto& \int \delta(H_B - E_B) \,\dbx_B\label{portos}
\eeqa
is the pdf of the weak vortex energy $H_B$ in the uniform distribution.
The last term in \Ref{sortos} means that the function $p_0(E_B)$ should be
evaluated at $E_B=E-H_A$.  Clearly, all states $\bx_A$ with the same energy
$H_A$ are now equally likely.  Because $\rho_E(\bx_A)$ depends only on
$E-H_A$ this is a huge simplification, as $p_0(E_B)$ can be computed
once and for all from the uniform distribution.

However, the Hamiltonian \Ref{HFCC} does not fit into this category, as we
have
\beql{HFtrouble}
H(\bx_A,\bx_B)  = H_A(\bx_A)+H_B(\bx_B) + H_I(\bx_A,\bx_B)
\eeq
where the `interfacial' energy $H_I$ consists of terms involving both strong
and weak vortex circulations.  In principle, this means that $\rho_E(\bx_A)$
does not depend solely on $E-H_A$.  Calculating $\rho_E(\bx_A)$ for all
$\bx_A$ directly would be very expensive, as even in the present case this
would require a look-up table in eight dimensions.  Instead, a much simpler
{\sl ad hoc\/} approximation for $\rho_E(\bx_A)$ as a function of $E-H_A$ is
used, which for $H_I=0$ reduces to \Ref{sortos}.  The approximation is
\beql{rhoarg}
\rho_E(\bx_A) \propto p_0(E_R=E-H_A(\bx_A)), 
\eeq
where 
\beql{p0er}
\obfbox{p_0(E_R) \propto \int\int \delta(H_R-E_R)\,\bdx_A\bdx_B}
\eeq
is the pdf of the `reservoir' energy $H_R\equiv H_B+H_I$ based on
the uniform distribution.  Unlike in \Ref{portos}, the double
integral is necessary here because $H_R$ depends on both $\bx_A$
and $\bx_B$.  This approximation effectively assigns the same
probabilitiy to all states $\bx_A$ that have the same strong
vortex energy $H_A(\bx_A)$.  Indeed, the assigned probability can
be shown to be the average probability over all states with the
same $H_A$.

Based on this approximation, the pdf for any $\Phi(\bx_A)$ is now
given by
\beql{lord1}
\obfbox{p_E(\phi) \propto \int \delta(\Phi-\phi)p_0(E_R=E-H_A)\,\dbx_A}.
\eeq
The pdf of $H_A$ in particular simplifies further to
\beql{pods}
p_E(E_A) \propto p_0(E_A)p_0(E_R=E-E_A).
\eeq

The additional consideration of the second (non-robust) angular momentum
invariant \Ref{AngMomDef} is straightforward, especially as $\hat M = \hat M_A +
\hat M_B$ holds exactly.  The microcanonical density becomes
\beql{rhomcdef}
\rho_{EM} = \frac{\delta(H-E)\delta(\hat M - M)}{\int \delta(H-E)\delta(\hat M
- M)\,\dbx^N}, 
\eeq
the marginal density 
\beql{barney}
\rho_{EM}(\bx_A)  \propto \int \delta(H-E)\delta(\hat M
- M)\,\dbx_B,
\eeq
and the joint `reservoir' pdf based on the uniform distribution
\beql{pquark}
\obfbox{p_0(E_R,M_B) \propto \int\int \delta(H_R-E_R)\delta(\hat M_B -
M_B)\,\bdx_A\bdx_B}.
\eeq
  The same approximation procedure for \Ref{barney} as before then gives
\beql{pdiag}
\obfbox{p_{EM}(\phi) \propto \int
\delta(\Phi-\phi)p_0(E_R=E-H_A,M_B=M-\hat M_A)\,\dbx_A}
\eeq
The predictions of the statistical mechanics theory are hence pdfs estimated
based on \Ref{pquark} and \Ref{pdiag}.  For comparison with \fref{fHamAm} the
functions $p_0(E_R)$ and $p_0(E_R,M_B)$ are plotted in \fref{fResHamAm}.

Despite the somewhat opulent appearance of \Ref{pquark} and \Ref{pdiag}, the
practical procedure for estimating pdfs is actually disarmingly simple.  A
sample of $10^5$ states was generated using the uniform distribution and the
corresponding values of $H,H_A,H_R,\hat M,\hat M_A,\hat M_B$ were computed and
stored in lists, as were the coordinates $\bx_A$ of the four strong vortices.
Computing $H$ is by far the most expensive step here.  Histograms based on
these lists were then used to estimate \Ref{pquark}.  For any function
$\Phi(\bx_A)$ to be investigated a corresponding list of values was then
computed from the stored coordinates.  This list together with a look-up table
for the histogram increments $\propto p_0(E_R,M_B)$ at the shifted arguments
was then used to estimate \Ref{pdiag} at fixed $E$ and $M$.  It is
worth stressing that only a single large sample based on the uniform
distribution is needed to describe the statistics of the system at arbitrary
$E$ and $M$.
\epsherel{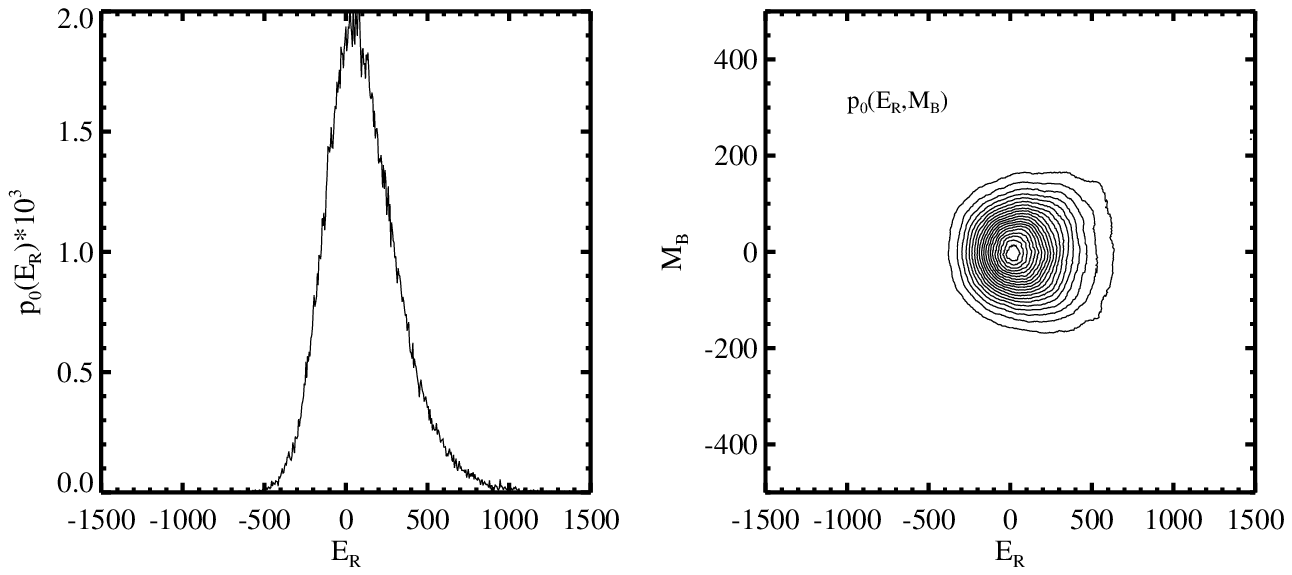}{(a) Pdf of the reservoir energy $E_R=E-E_A$, where
$E_A$ is the energy of the strong vortices, based on $10^5$ random samples with
uniform distribution.  The average reservoir energy is $109$ with variance
$229^2$.  (b) Joint reservoir energy--angular momentum pdf, after some
smoothing, where $M_B=M-M_A$ and $M_A$ is the angular momentum of the strong
vortices.}{fResHamAm}

\subsection{Comparison with model results }

The pdfs for $E_A$ (i.e.\ the energy of the strong vortices) are plotted in
\fref{fHamPdfs}.
\epsherel{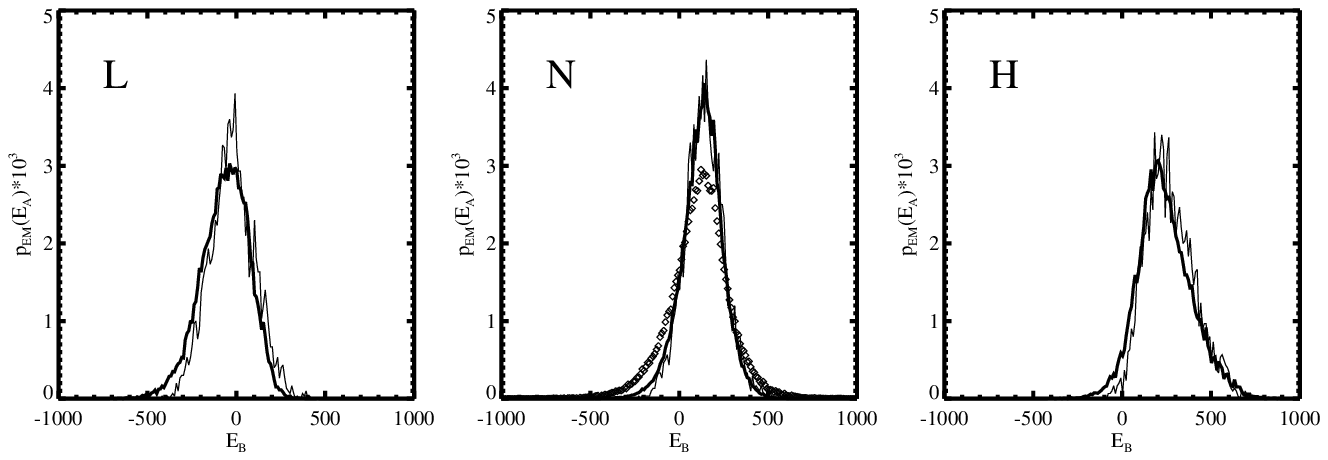}{Pdfs of $E_A$, the energy of the strong vortices, for the
three runs.  Thin lines are simulation results, thick lines are theoretical 
predictions, and the squares denote $p_0(E_A)$.}{fHamPdfs}
Throughout, thin lines denote pdfs estimated from histograms taken from the
direct numerical simulations and thick lines denote theoretical predictions
based on \Ref{pdiag}.  Also shown in the middle panel is $p_0(E_A)$, which
corresponds to random placing of the vortices with uniform distribution.  As
could be expected, in the neutral energy case this gives a reasonable first
approximation, though still a less accurate one than $p_{EM}(E_A)$.  The mean
energy of the strong vortices increases as $E$ increases and this can clearly
be predicted quantitatively from the theoretical predictions.  In most cases,
there is very good agreement between the theoretical and simulation
statistics, not only in terms of accurate prediction of low-order moments, but
also in the prediction of the non-Gaussian shape of the pdfs.

Figures~\ref{fDistSPdfs}-\ref{fDistDPdfs} show the pdfs for the distance
between same- and opposite-signed strong vortices, respectively.
\epsherel{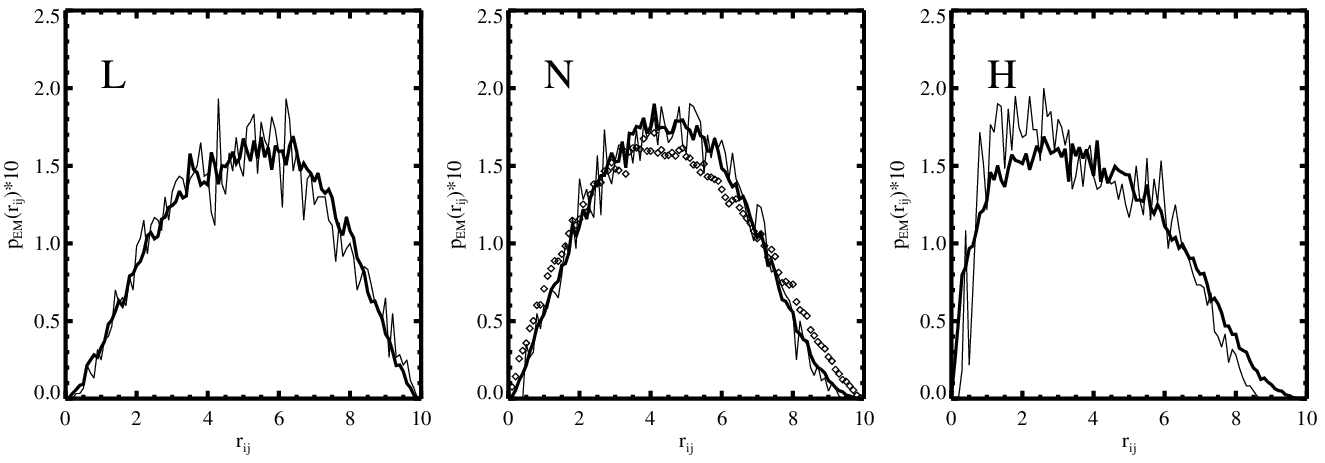}{Pdfs of $r_{ij}$ between strong vortices of the same
sign, for the three runs.  Thin lines are simulation results, thick lines are
theoretical predictions, and the squares denote $p_0(r_{ij})$.  The simulation
results are averages over the pdfs of $r_{12}$ and $r_{34}$.}{fDistSPdfs}
\epsherel{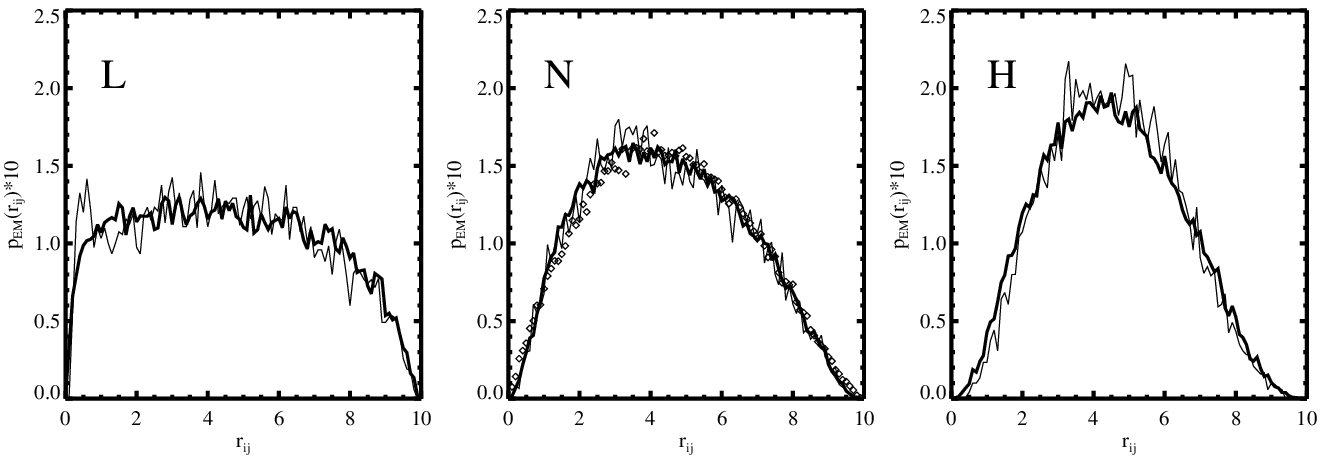}{Pdfs of $r_{ij}$ between strong vortices of opposite
sign, for the three runs.  Thin lines are simulation results, thick lines are
theoretical predictions, and the squares denote $p_0(r_{ij})$.  The simulation
results are averages over the pdfs of $r_{13},r_{14},r_{23},$ and
$r_{24}$.}{fDistDPdfs}  
The tendency for same-signed vortices to cluster at smaller $r_{ij}$ with
increasing $E$ is evident in \fref{fDistSPdfs}.  On the other hand, the
indifference to $E$ of the average $r_{ij}$ between opposite-signed vortices
that was noted in table~\ref{tmom} masks notable changes in the pdf that are
evident in \fref{fDistDPdfs}.  These are well captured by the theoretical
predictions.

Finally, \fref{fRadPdfs} shows statistics for $r_i$, the vortex distance from
the origin.  This quantity is interesting due to the influence of the
self-interaction terms in \Ref{HFCC}, which, as noted in \S\ref{hamsec},
significantly affect the dynamics of the strong vortices.
\epsherel{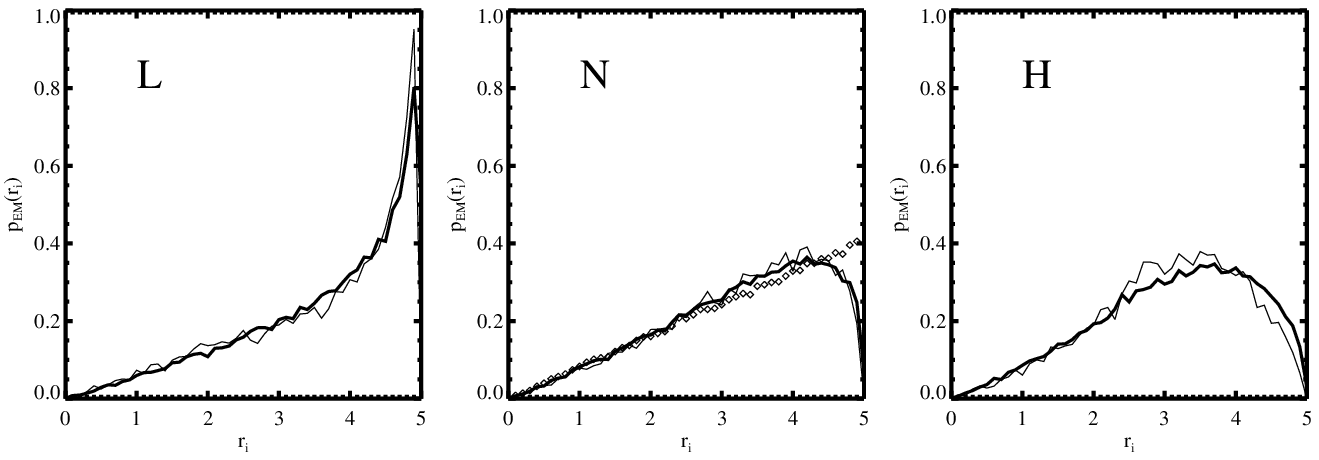}{Pdfs of $r_i$, the distance from the origin of the
strong vortices, for the three runs.  Thin lines are simulation results, thick
lines are theoretical predictions, and the squares denote $p_0(r_{i})$.  The
simulation results are averages over the pdfs of $r_{1},r_{2},r_{3},$ and
$r_{4}$.}{fRadPdfs}
For neutral and high energies the pdf of $r_i$ settles down to a shape quite
close to the uniform distribution (except near the wall $r_i=R$), which is
indicated by the squares in the middle panel. However, the first panel in
\fref{fRadPdfs} shows how the strong vortices tend to accumulate at the
cylinder wall for very low energies.  This is essentially a similar
statistical effect as the formation of opposite-signed vortex dipoles in the
first panel of \fref{fDistDPdfs}, but the wall effect is clearly more
pronounced in its pdf.  Interestingly, the effective temperature estimated as
d$\ln p_0(E)$/d$E$ from \fref{fHamAm} at $E=-197$ gives $\approx0.01$ for this
run.  The theoretical upper limit in \Ref{betarange2} for the possible
existence of canonical subsystem statistics in the present case gives
$\approx0.013$, which confirms that this run is close to a collapse to the
wall.  Finite point vortex statistics in a collapse case would rely entirely
on the finite size of the reservoir formed by the small vortices.

%
%
\section{CONCLUDING REMARKS}\label{concsec}

The theoretical predictions based on an ergodic approximation for the whole
vortex system were seen to predict the pdfs of many, though not all,
descriptive variables of the strong vortex subsystem with a surprisingly high
degree of accuracy.  Some simulation averages converged only very slowly and
significantly longer integrations could test the quality of the ergodic
approximation in these cases.  On the other hand, non-ergodic observations
might also be due to the crucial approximation leading to \Ref{pdiag}, which
was necessary to estimate the theoretical pdfs at affordable computational
cost.  Otherwise, the prediction procedure was remarkably simple and cheap,
relying only on a single random sample of vortex configurations to predict
pdfs for {\em all\/} values of total energy $E$ and angular momentum $M$.

The toy model, the direct numerical vortex simulations, and the theoretical
predictions all corroborated Onsager's crucial insight that strong vortices
will exhibit amplified statistical tendencies compared to weak
vortices, and hence will be more predictable in a negative temperature state.
It is the strongly unequal circulation strengths that allows the flow to
organise itself in this inhomogeneous manner.  It can be noted that in terms
of entropy as a measure of accessible phase space volume, clustered vortices
always present a low-entropy state.  The crucial point is that in a negative
temperature state the low entropy of the clustering strong vortices is more
than compensated for by the high entropy of the freely roaming weak vortices.

It is intriguing to note that on the level of individual vortex
dynamics there is a smooth transition from positive to negative
temperature behaviour, exemplified by the smooth $p_0(E)$ in
\fref{fHamAm}.  By contrast, in a coarse-grained picture two
vortices with opposite circulations that are close together
cancel each other out and hence disappear from view.  This
illustrates why solutions to mean-field theories (such as the
Sinh--Poisson equation$^4$) have a characteristic cut-on
behaviour as $\beta<0$, because only same-signed vortex
clustering is observable in the coarse-grained variables of these
theories.  It is also noteworthy that the near-collapse of the
vortices to the wall in the present low-energy case (which is
linked to the theoretical upper bound in \Ref{betarange2}) does
not occur in the Sinh--Poisson equation because there the scaling
has been arranged {\it ab initio\/} to render the wall-induced
self-interaction energies negligible.

The present set-up seems to have been the most complicated to
study, i.e.\ the strong vortices interact vigorously with both
the wall and with the small vortices, and the small vortices
themselves form only a finite energy and angular momentum
reservoir.  As noted before, the latter point puts the strong
vortex statistics beyond the reach of the usual canonical
theories.  In other words, whilst the overall energy regime can
be broadly classified by the sign of the usual statistical
mechanics temperature, the temperature concept alone is not
sufficient to make quantitative predictions.  The modified
maximum entropy principle suggested by the Gaussian pdfs of the
toy model (in which entropy was maximized subject to both a fixed
mean energy and a fixed energy variance) might allow analytical
progress to be made here.  Another possible avenue for future
analytical progress is an asymptotic exploitation of the small
parameter $|\Gamma_B/\Gamma_A|$.

It is tempting to generalize the present results (which directly
apply only to well-separated finite-size vortices) to continuous
vorticity distributions by letting $N\to\infty$ in some way.  As
is now well known, the relevant scaling must keep $N\Gamma=O(1)$ in
this limit.  However, this implies that the microscopic vortex
mobility due to the induced velocity by the nearest vortices at
average distance $\propto R/\sqrt{N}$ would decrease to zero as
$\Gamma\sqrt{N}\propto1/\sqrt{N}$.  Therefore the microscopic
vortex system equilibration time (which is implicitly assumed
here to be small compared to the observation period) goes to
infinity in this limit.  This problem is directly observable,
say, in the simplest case of many $\Gamma_i=$~const.\ vortices in
a cylinder, for which exact solutions of the $N\to\infty$ limit
are known$^8$.  Simulations starting from non-equilibrium initial
conditions clearly show that these solutions are practically
unattainable due the lack of vortex mobility.  On the other hand,
there is evidence$^3$ that the slow evolution of well-mixing
large-scale flows can be approximated by statistical point vortex
theory.  Notably, random forcing also helps in this context, as
it too increases the mobility of the vortex population.  This
suggests that a limit in which $N_A=$~const.\ and $N_B\to\infty$
could perhaps be used to model the behaviour of strong vortices
surround by a sea of ``filamentary vortex debris'', because the
stirring by the strong vortices could provide the essential
mobility for the debris.

Finally, the prediction tools developed in this paper could be
used to study an interesting `inverse' problem: from observing
only the strong vortices, can one deduce the number and strengths
of the unobserved weak vortices?  This would provide a nonlinear
method for estimating unobservable sub-gridscale data, perhaps
with applications in geophysical fluid dynamics.

ACKNOWLEDGEMENTS

The idea for the work grew out of a project undertaken at the
2000 summer study in Geophysical Fluid Dynamics hosted by the
Woods Hole Oceanographic Institution (USA) and sponsored by the
US~National Science Foundation.  Rick Salmon, the director of the
2000 summer study, is thanked for his support.  Further financial
support from the Nuffield Foundation~(UK) under grant NAL/00034/G
and from the Engineering and Physical Sciences Research
Council~(UK) under grant GR/R09565/01 is gratefully acknowledged.
Ben Aschenbrenner is thanked for performing some early
simulations.

\newpage
\centerline{\sc References}

\paper $^1$Onsager, L., 1949:
Statistical hydrodynamics.
Nuovo Cimento (suppl.),
6, 279--287.

\paper $^2$Kraichnan, R.\ H., Montgomery, D., 1980:
Two-dimensional turbulence.
Rep.\ Prog.\ Phys.,
43, 547--619.

\paper $^3$Grote, M.\ J., Majda, A.\ J., 1997:
Crude closure dynamics through large scale statistical theories.
Phys.\ Fluids,
9, 3431--3442.

\paper $^4$Joyce, G., Montgomery, D., 1973:
Negative temperature states for the two-dimensional guiding-centre plasma.
J.\ Plasma Phys.,
10, 107--121.

\paper $^5$Montgomery, D., Joyce, G., 1974:
Statistical mechanics of ``negative temperature'' states.
Phys.\ Fluids,
17, 1139--1145.

\paper $^6$Pointin, Y.\ B., Lundgren, T.\ S., 1976:
Statistical mechanics of two-dimensional vortices in a bounded container.
Phys.\ Fluids,
19, 1459--1470.

\paper $^7$Eyink, G.\ L., Spohn, H., 1993:
Negative-temperature states and large-scale, long-lived vortices in
two-dimensional turbulence.
J.\ Stat.\ Phys.,
70, 833--886.

\paper $^8$Caglioti, E., Lions, P.\ L., Marchioro, C., Pulvirenti, M., 1992:
A special class of stationary flows for two-dimensional Euler equations: a
statistical mechanics description.
Comm.\ Math.\ Phys.,
143, 501--525.

\paper $^9$Lions, P.\ L., Majda, A.\ J., 2000:
Equilibrium statistical theory for nearly parallel vortex filaments.
Comm.\ Pure App.\ Math.,
8, 76--142.

\paper $^{10}$Weiss, J.\ B., McWilliams, J.\ C., 1991:
Nonergodicity of point vortices.
Phys.\ Fluids A,
3, 835--844.

\paper $^{11}$Weiss, J.\ B., Provenzale, A., McWilliams, J.\ C., 1998:
Lagrangian dynamics in high-dimensional point-vortex systems.
Phys.\ Fluids,
10, 1929--1941.

\book $^{12}$Lamb, H., 1932:
Hydrodynamics, 6th ed.
Cambridge University Press,
738pp.

\paper $^{13}$B\"uhler, O., Jacobson, T.\ E., 2001:
Wave-driven currents and vortex dynamics on barred beaches.
J.\ Fluid Mech., 449, 313--339.

\book $^{14}$Chorin, A.\ J., 1994:
Vorticity and turbulence.
Springer,
174pp.

\end{document}